\documentclass{iaus}
\usepackage{graphicx}

\title[Helicity transport in a simulated CME] 
      {Helicity transport in a simulated coronal mass ejection}

\author[B.~Kliem et al.]   
       {B.~Kliem$^{1,2,3}$,
        S.~Rust$^1$ \and 
        N.~Seehafer$^1$}

\affiliation{
 $^1$Institute of Physics and Astronomy, University of Potsdam,
     Karl-Liebknecht-Str.\ 24-25, 14476~Potsdam, Germany\\[\affilskip] 
 $^2$Mullard Space Science Laboratory, University College London,
     Holmbury St.\ Mary, Dorking, Surrey, RH5 6NT, UK\\[\affilskip] 
 $^3$Space Science Division, Naval Research Laboratory,
     Washington, DC 20375, USA}

\pubyear{2011}
\volume{274}  
\pagerange{1--2}
\jname{Advances in Plasma Astrophysics}
\editors{A.~Bonanno, E.~de~Gouveia~dal~Pino \& A.~Kosovichev, eds.}
\begin{document}

\maketitle

\begin{abstract} It has been suggested that coronal mass ejections
(CMEs) remove the magnetic helicity of their coronal source region from
the Sun. Such removal is often regarded to be necessary due to the
hemispheric sign preference of the helicity, which inhibits a simple
annihilation by reconnection between volumes of opposite chirality. Here
we monitor the relative magnetic helicity contained in the coronal
volume of a simulated flux rope CME, as well as the upward flux of
relative helicity through horizontal planes in the simulation box. The
unstable and erupting flux rope carries away only a minor part of the
initial relative helicity; the major part remains in the volume. This is
a consequence of the requirement that the current through an expanding
loop must decrease if the magnetic energy of the configuration is to
decrease as the loop rises, to provide the kinetic energy of the CME.

\end{abstract}

\firstsection 
\section{Introduction}
\label{introduction}

The helicity of the solar magnetic field obeys a hemispheric preference
which is invariant with respect to the sign reversal of the global
magnetic field with the activity cycle (\cite[Hale 1925]{Hale1925};
\cite[Seehafer 1990]{Seehafer1990}). This has led to the suggestion that
coronal mass ejections (CMEs) must remove magnetic helicity from the Sun
to prevent indefinite accumulation of the helicity in each hemisphere
(\cite[Rust 1994]{Rust1994}; \cite[Low 1996]{Low1996}). It was also
found that the accumulation and removal of helicity can control the rate
of mean-field dynamo action, so that the evolution of the activity cycle
may be related to the flow of helicity through the Sun (\cite[Blackman
\& Field  2001]{Blackman&Field2001}; \cite[Brandenburg \& Subramanian
2005]{Brandenburg&Subramanian2005}). Careful studies of the long-term
helicity budget of two solar active regions (\cite[D{\'e}moulin et al.
2002]{Demoulin&al2002}; \cite[Green et al. 2002]{Green&al2002}) appear
to confirm the conjecture of efficient helicity shedding by CMEs.
However, both investigations used the linear force-free field
approximation to estimate the helicity in the active region atmosphere.
The accuracy of this estimate is not known. Similarly, the estimates of
the helicity in interplanetary CMEs are still subject to considerable
uncertainty (\cite[Demoulin 2007]{Demoulin2007}). In this paper,
numerical simulation is used to quantify the transport of magnetic
helicity from the source volume of CMEs.

\section{Relative magnetic helicity in a simulated CME}
\label{helicity}

\begin{figure}
 \centering
 \includegraphics[width=.9\textwidth]{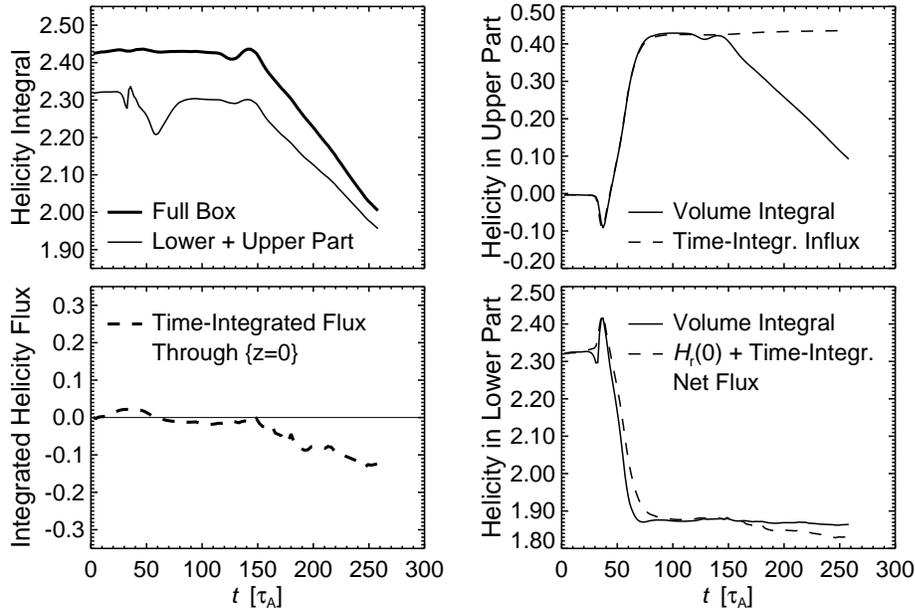}  
 \caption{
  Relative magnetic helicity $H_\mathrm{r}(t)$ in the simulation box and
  in the lower and upper sub-volumes, and time-integrated helicity
  fluxes trough the bottom and diagnostics planes (bottom right) and
  through the diagnostics plane only (top right) in a CME simulation.
  The erupting flux begins to cross the diagnostics plane and the upper
  boundary of the box at $t\approx30\tau_\mathrm{A}$ and
  $t\approx120\tau_\mathrm{A}$, respectively. A helicity of about 0.4
  (in normalized units) is transported from the lower into the upper
  sub-volume and then out of the box. This is about 1/5 of the initial
  helicity $H_\mathrm{r}(0)\approx2.4$.
  (The minor outflux of helicity through the bottom boundary for
  $t\gtrsim150\tau_\mathrm{A}$ results from downward propagating
  perturbations triggered by the ejected flux at large heights and may
  be a numerical artefact.
  The simulation was terminated by numerical instability in one of the
  upper corners of the box, where an open and two closed boundaries meet
  and numerical stability is more difficult to maintain than in the
  interior of the box.)}
 \label{f:H(t)}
\end{figure}

We monitor the relative magnetic helicity
$H_\mathrm{r}=\int({\bf A}+{\bf A}_\mathrm{p})\cdot
                  ({\bf B}-{\bf B}_\mathrm{p})\,\mathrm{d}V$
(\cite[Berger \& Field 1984]{Berger&Field1984}; \cite[Finn \& Antonsen
1985]{Finn&Antonsen1985}) in an MHD simulation of a flux rope CME.  The
force-free equlibrium of a toroidal current channel partially submerged
below the photosphere (\cite[Titov \& D{\'e}moulin
1999]{Titov&Demoulin1999}) is chosen as initial condition. The twist is
set to a supercritical value, so that the kink-unstable magnetic flux
rope formed by the current channel spontaneously starts to rise. The
ideal MHD equations are integrated, neglecting pressure, with numerical
diffusion enabling magnetic reconnection. The simulation is similar to
the one in \cite[T{\"o}r{\"o}k \& Kliem (2005)]{Torok&Kliem2005}, except
for an open upper boundary and a larger box size (of $40^3$ unit
lengths, set to be the initial flux rope apex height $h_0$).
Approximations of the instantaneous helicity in the simulation box and
of the helicity flux through the bottom boundary are obtained from
Equations~(1)--(7) in \cite[DeVore (2000)]{DeVore2000}. These are exact
only if the field strength has fallen to zero at the top and lateral
boundaries of the considered volume. We have checked that the computed
approximate helicity of the initial equilibrium approaches a limit for
increasing box size and that it deviates by less then 2\% from the
apparent limit value for the chosen size.

The approximation degrades as the flux rope passes through the top
boundary, although the field strength then still decreases by more than
two orders of magnitude from the bottom to the top of the box. To
estimate the magnitude of the error, we divide the box by a horizontal
``diagnostics plane'' at height $z=6h_0$, such that the field drops with
increasing height by a similar factor in each sub-volume. At $t=0$ the
factor is of order $10^2$, justifying the use of the approximation in
each sub-volume. The top left panel in Figure~\ref{f:H(t)} shows that
the summed helicities of the sub-volumes differ by $\lesssim5\%$ from
the helicity of the box as a whole, with the error obviously originating
from the lower sub-volume, since $H_\mathrm{r}(t\!=\!0)=0$ is correctly
found in the upper sub-volume (top right panel). When the upper part of
the unstable flux rope begins to propagate through the diagnostics
plane, the approximation is degraded in the lower sub-volume, but is
still of similar quality as before in the upper sub-volume and in the
box as a whole. The resulting error is given by the additional
difference between the two values for the whole volume in the relevant
time interval, $t\sim(30\mbox{--}80)\tau_\mathrm{A}$, where
$\tau_\mathrm{A}=h_0/V_\mathrm{A}$ and $V_\mathrm{A}$ is the initial
Alfv\'en velocity in the flux rope. The error reaches a peak value
$\lesssim5\%$ at $t=60\tau_\mathrm{A}$, when the flux rope apex has
risen to $z=17h_0$, and decreases considerably thereafter. The helicity
calculation for $t\gtrsim120\tau_\mathrm{A}$, when the CME leaves the
box, should even be more precise, since the field strengths of the flux
rope at the upper boundary are then smaller, by an order of magnitude,
than the field strengths at the passage of the diagnostics plane.

Figure~\ref{f:H(t)} shows that a relative helicity of $\approx\!0.4$ (in
the normalized units of the simulation, which prescribe a field strength
of unity at the apex of the initial flux rope, $|{\bf B}_0(0,0,1)|=1$)
is transported from the lower into the upper subvolume and then out of
the box when the upper part of the erupting flux rope crosses the
respective top boundary. (A helicity of $\approx\!0.1$ is not yet
ejected when the simulation is terminated due to numerical instability,
but it is clear from the plot that this will be ejected as well.) The
top right plot shows a very small rate of helicity flux through the
diagnostics plane after the top part of the flux rope has propagated
into the upper sub-volume ($t\gtrsim80\tau_\mathrm{A}$). This suggests
that the further upward stretching of the flux rope legs and the
addition of flux to the rope by reonnection in the vertical current
sheet under the rope do not contribute strongly to the ejected helicity,
at least not at the scales covered by the simulation, which extend to
flux rope apex heights of $\approx\!25$ times the footpoint distance
(several solar radii when scaled to a large solar active region). The
upward reconnection outflow velocity, a proxy of the reconnection rate
in our ideal MHD simulation, has decreased to one quarter of its peak
value, $u_z(t=52)=0.8V_\mathrm{A}$, by the end of the simulation. The
soft X-ray flux of long-duration solar ejective events, an indicator of
the energy release by reconnection, decreases strongly from its peak
value on such scales. Therefore, although this simulation is terminated
by numerical instability, it likely models most of the helicity ejection
in a CME. However, this amounts to only about 1/5 of the initial
relative helicity.

\begin{figure}
 \centering
 \includegraphics[width=.63\textwidth]{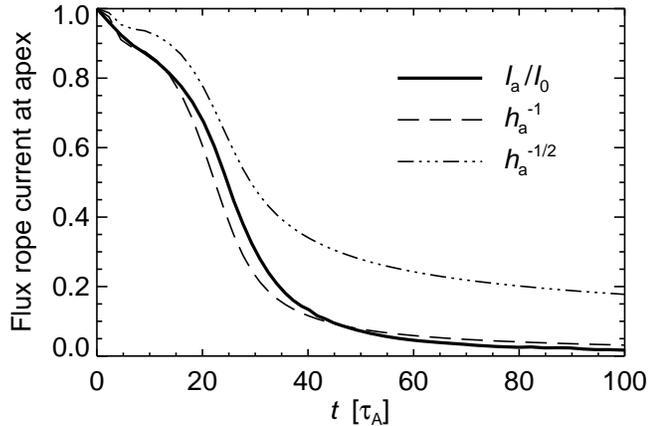}  
 \caption{
  Total current through the apex of the flux rope, normalized by initial
  current, $I_\mathrm{a}(t)/I_0$. For comparison, the inverse of the
  apex height $h_\mathrm{a}(t)$, a proxy for the flux rope length, and
  $h_\mathrm{a}(t)^{-1/2}$ are shown (see text).}
 \label{f:I(t)}
\end{figure}

The relatively small efficiency of helicity ejection can be related to
the evolution of the current distribution in the volume. These currents
carry the relative helicity (which vanishes in a potential field). The
initial coronal equilibrium consists of a section of a toroidal current
ring. The energy of a current ring is given by $W=LI^2/2\approx
I^2R\,[\ln(8R/a)-7/4]$, where $I$, $R$, $a$, and $L$ are the current,
major and minor radius, and the inductance of the ring, respectively.
Since magnetic energy must be released in order to accelerate the ejecta
and the term in brackets does not vary strongly, the current in the
rising flux loop of the CME must decrease faster than $R^{-1/2}$. In the
approximation of ideal MHD the current in the loop decreases roughly as
$R^{-1}$, because the number of field line turns in the loop is
conserved. The simulation shows such a fast decrease
(Figure~\ref{f:I(t)}), which results in most of the current staying low
in the box. Consequently, only a minor part of the initial relative
helicity leaves the system with the ejected flux.

\section{Conclusions and Discussion}
\label{conclusions}

The simulated CME ejects only a minor part of the initial relative
magnetic helicity from its source volume. Although this result requires
substantiation through the study of its parametric dependence and of
other equilibria, the necessary decrease of the current through an
expanding unstable flux loop leads us to expect that it holds generally.
The number of CMEs per active region varies within very wide limits.
Between 30 and 65 CMEs have been estimated to occur throughout the
lifetime of the two very CME-prolific active regions studied in
\cite[D{\'e}moulin et al.\ (2002)]{Demoulin&al2002} and \cite[Green et
al.\ (2002)]{Green&al2002}. On the other hand, the majority of active
regions produces no CME at all, or only one CME in their lifetime.
Hence, the shedding of helicity by CMEs may be of lower importance than
originally conjectured.

It appears natural to assume that, perhaps generally, much of an active
region's helicity submerges when the region disperses and the major part
of its flux submerges below the photosphere. The helicity may then
follow the slow journey of magnetic flux in the course of the solar
cycle. Annihilation of helicity in the interior of the Sun, following
the transport of the helicity-carrying flux to the equatorial plane by
the meridional flow, is one possibility to prevent the helicity in each
hemisphere from accumulating indefinitely. Another possibility, opposite
to a common conjecture, is that the helicity in the solar interior, like
magnetic energy, undergoes a normal (or direct) turbulent cascade
towards small spatial scales, where it is dissipated. Also, the cascade
directions may be different for small-scale and large-scale fields
(\cite[Alexakis et al.\ 2006]{Alexakis&al2006}), with the helicity of
active-region magnetic fields, considered to be small-scale fields,
subject to a direct cascade.


\end{document}